\documentclass{myllncs-mixalis}
\usepackage{graphicx}
\usepackage{amsmath}
\usepackage{amssymb}
\usepackage[ruled, noline, algo2e, noend]{algorithm2e}
\usepackage[T1]{fontenc}
\usepackage{enumerate}
\usepackage[font=small, format=hang, labelfont=bf]{caption}
\usepackage{subfig}
\usepackage{color}

\SetArgSty{}  \SetKw{KwOr}{or}
\SetKw{KwAnd}{and} \SetKw{KwInvariant}{invariant:}
\SetKw{KwDownTo}{down to} \SetKwInOut{Input}{input}
\SetKwInOut{Output}{output} \SetKwInOut{Require}{require}
\SetKwComment{Comment}{start}{end}
\SetKwIF{If}{ElseIf}{Else}{if}{then}{else if}{else}{endif}
\dontprintsemicolon

\newcommand{\eat}[1] {{}}

\begin{document}
\parindent=0cm
\parskip=5pt

\title{Optimal Acyclic Hamiltonian Path Completion for Outerplanar Triangulated
$st$-Digraphs\\
(with Application to Upward Topological Book Embeddings)}
\author{
Tamara  Mchedlidze 
,  Antonios Symvonis 
}

\institute{%
    Dept. of Mathematics, National Technical University of Athens,
 Athens, Greece.\\
    \texttt{\{mchet,symvonis\}@math.ntua.gr}
}

\maketitle

\vspace{-10pt}

\begin{abstract}

Given an embedded  planar acyclic digraph $G$, we define the problem
of \emph{acyclic hamiltonian path completion with crossing
minimization (Acyclic-HPCCM)} to be the problem of determining an
\emph{hamiltonian path completion set} of edges such that, when
these edges are embedded on $G$, they create the smallest possible
number of edge crossings and turn  $G$ to a hamiltonian digraph. Our
results include:
\begin{enumerate}
\item
We provide a characterization under which a triangulated
$st$-digraph $G$ is hamiltonian.
\item For an outerplanar triangulated $st$-digraph $G$,
we define the \emph{$st$-polygon decomposition of $G$} and, based on
its properties, we develop a linear-time algorithm that solves the
Acyclic-HPCCM problem with at most one crossing per edge of $G$.
\item
For the class of $st$-planar digraphs, we establish an equivalence
between the Acyclic-HPCCM problem and the problem of determining an
upward 2-page topological book embedding with  minimum number of
spine crossings. We infer (based on this equivalence) for the class
of outerplanar triangulated $st$-digraphs an upward topological
2-page book embedding with minimum number of spine crossings and at
most one spine crossing per edge.
\end{enumerate}

To the best of our knowledge, it is the first time  that
edge-crossing minimization is studied in conjunction with the
acyclic hamiltonian completion problem and the first time that an
optimal algorithm with respect to spine crossing minimization is
presented for upward topological book embeddings.
\end{abstract}

\eat{
\textbf{TO FIX IN PAPER}
\begin{enumerate}
\item

\end{enumerate}
}


\section{Introduction}
In the \emph{hamiltonian path completion problem} (for short,
\emph{HP-completion}) we are given a graph $G$ (directed or
undirected) and we are asked to identify a set of edges (refereed to
as an \emph{HP-completion set}) such that, when these edges are
embedded on  $G$ they turn it to a hamiltonian graph. The resulting
hamiltonian graph $G^\prime$ is referred to as the
\emph{HP-completed graph} of $G$.
 When we treat the HP-completion problem as an
optimization problem, we are interested on an HP-completion set of
minimum size.

When the input graph $G$ is a planar embedded digraph, an
HP-completion set for $G$ must be naturally extended to include an
embedding of its edges on the plane, yielding to an embedded
HP-completed digraph $G^\prime$. In general, $G^\prime$ is not
planar, and thus, it is natural to attempt to minimize the number of
edge crossings of the embedding of the HP-completed digraph
$G^\prime$ instead of the size of the HP-completion set. We refer to
this problem as the \emph{HP-completion with crossing minimization
problem} (for short, \emph{HPCCM}). The HPCCM problem can be further
refined  by placing restrictions on the maximum number of permitted
crossings per edge of  $G$.

When the input digraph $G$ is acyclic, we can insist on
HP-completion sets which leave the HP-completed digraph $G^\prime$
also acyclic. We refer to this version of the problem as the
\emph{acyclic HP-completion problem}.

A \emph{k-page book} is a structure consisting of a line, referred
to as \emph{spine}, and of $k$ half-planes, referred to as
\emph{pages}, that have the spine as their common boundary. A
\emph{book embedding} of a  graph $G$ is a drawing of $G$ on a book
such that the vertices are aligned along the spine, each edge is
entirely drawn on a single page, and edges do not cross each other.
If we are interested only on two-dimensional structures we have to
concentrate on  2-page book embeddings and to allow spine crossings.
These embeddings are  also referred to as  2-page\emph{ topological
}book embeddings.

For acyclic digraphs, an upward book embedding can be considered to
be a book embedding in which the spine is vertical and all edges are
drawn monotonically increasing in the upward direction. As a
consequence, in an upward book embedding of an acyclic digraph the
vertices appear along the spine in topological order.

The results on topological book embedding that appear in the
literature focus on the number of spine crossings per edge required
to book-embed a graph on a 2-page book. However, approaching the
topological book embedding problem as an optimization problem, it
make sense to also try to minimize the number of spine crossings.

In this paper, we introduce the problem of \emph{acyclic hamiltonian
path completion with crossing minimization} (for short,
\emph{Acyclic-HPCCM}) for planar embedded acyclic digraphs. To the
best of our knowledge, this is the first time that edge-crossing
minimization is studied in conjunction with the acyclic
HP-completion problem. Then, we provide a characterization under
which a triangulated $st$-digraph is hamiltonian. For an outerplanar
triangulated $st$-digraph $G$, we define the \emph{$st$-polygon
decomposition of $G$} and, based on the decomposition's properties,
we develop a linear-time algorithm that solves the Acyclic-HPCCM
problem with at most one crossing per edge of $G$.

In addition, for the class of $st$-planar digraphs, we establish an
equivalence between the acyclic-HPCCM problem and the problem of
determining an upward 2-page topological book embeddig with a
minimal number of spine crossings. Based on this equivalence, we can
infer  for the class of outerplanar triangulated $st$-digraphs an
upward topological 2-page book embedding with minimum number of
spine crossings and at most one spine crossing per edge. Again, to
the best of our knowledge, this is the first time that an optimal
algorithm with respect to spine crossing minimization is presented
for upward topological book embeddings.

\subsection{Problem Definition}
\label{sec:problemDefinition}
 Let $G=(V,E)$ be a  graph. Throughout
the paper, we use the term \emph{``graph''} we refer to both
directed and undirected graphs. We use the term ``digraph'' when we
want to restict our attention to directed graphs. We assume
familiarity with basic graph theory~\cite{Harary72,Diestel05}. A
\emph{hamiltonian path} of $G$ is a  path that visits every vertex
of $G$ exactly once. Determining whether a graph has a hamiltonian
path or circuit is NP-complete~\cite{GareyJS74}. The problem remains
NP-complete for cubic planar graphs~\cite{GareyJS74}, for maximal
planar graphs~\cite{Wigderson82}  and for planar
digraphs~\cite{GareyJS74}. It can be trivially solved in polynomial
time for planar acyclic digraphs.

Given a graph $G=(V,E)$, directed or undirected, a non-negative
integer $k \leq |V|$ and two vertices $s,~t \in V$, \emph{the
hamiltonian path completion (HPC)} problem asks whether there exists
a superset $E^\prime$ containing $E$ such that $|E^\prime- E| \leq
k$  and the graph $G^\prime = (V, E^\prime)$ has a hamiltonian path
from vertex $s$ to vertex $t$. We refer to  $G^\prime$ and to the
set of edges $|E^\prime- E|$ as the \emph{HP-completed graph} and
the \emph{HP-completion set} of graph $G$, respectively. We assume
that all edges of a HP-completion set are part of the Hamiltonian
path of $G^\prime$, otherwise they can be removed. When $G$ is a
directed acyclic graph, we can insist on HP-completion sets which
leave the HP-completed digraph also acyclic. We refer to this
version of the problem as the \emph{acyclic HP-completion problem}.
The hamiltonian path completion problem is
NP-complete~\cite{GareyJ79}. For
acyclic digraphs 
the  HPC problem is solved in polynomial time~\cite{KarejanM80}.

A \emph{drawing} $\Gamma$  of graph $G$ maps every vertex $v$ of $G$
to a distinct point $p(v)$ on the plane and each edge $e=(u,v)$ of
$G$ to a simple Jordan curve joining $p(u)$ with $p(v)$. A drawing
in which every edge $(u,v)$ is a a simple Jordan curve monotonically
increasing in the vertical direction is an \emph{upward  drawing}. A
drawing $\Gamma$ of graph $G$ is \emph{planar} if no two distinct
edges intersect except at their end-vertices. Graph $G$ is  called
\emph{planar} if it admits a planar drawing $\Gamma$.

An embedding of a planar graph $G$ is the equivalence class of
planar drawings of $G$ that define the same set of faces or,
equivalently, of face boundaries. A planar graph together with the
description of a set of faces $F$ is called an \emph{embedded planar
graph}.

Let $G=(V,E)$ be an embedded planar graph, $E^\prime$ be a superset
of   edges containing $E$, and $\Gamma(G^\prime)$ be a drawing of~
$G^\prime=(V, E^\prime)$. When the deletion from $\Gamma(G^\prime)$
of the edges in $E^\prime -E$ induces the embedded planar graph $G$,
we say that $\Gamma(G^\prime)$ \emph{preserves the embedded planar
graph $G$}.

\begin{definition}
\label{def:HPCCM} Given an embedded planar  graph   $G=(V,E)$,
directed or undirected, a non-negative integers $c$, and two
vertices $s,~t \in V$, the \emph{hamiltonian path completion with
edge crossing minimization (HPCCM) problem}  asks whether there
exists a superset $E^\prime$ containing $E$ and a drawing
$\Gamma(G^\prime)$ of graph $G^\prime = (V, E^\prime)$ such that (i)
$G^\prime$ has a hamiltonian path from vertex $s$ to vertex $t$,
(ii) $\Gamma(G^\prime)$ has at most $c$ edge crossings, and (iii)
$\Gamma(G^\prime)$ preserves the embedded planar graph $G$.
\end{definition}

We refer to the version of the HPCCM problem where the input is
directed acyclic graph and we
 insist on HP-completion
sets which leave the HP-completed digraph also acyclic as the
\emph{Acyclic-HPCCM} problem.

Over the set of all HP-completion sets for a graph $G$, and over all
of their different drawings that respect $G$, the one with a minimum
number of edge-crossings is called a \emph{crossing-optimal
HP-completion set}.

\begin{figure}[htb]
    \begin{minipage}{\textwidth}
    \centering
    \includegraphics[width=1\textwidth]{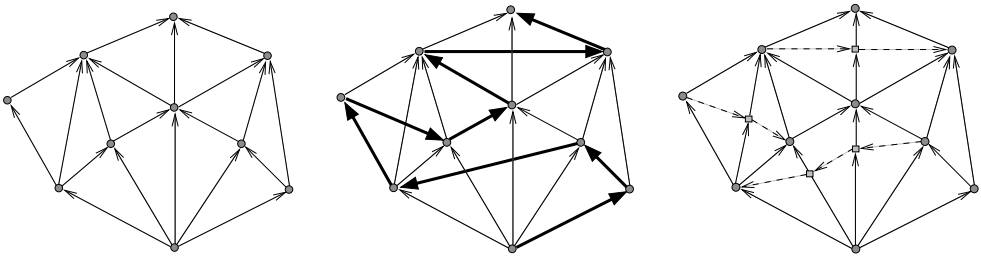}
    \caption{An embedded digraph $G$, a drawing $\Gamma$ of an HP-completed digraph of $G$
    and an HP-extended digraph of $G$ w.r.t. $\Gamma$. The edges of the HP-completion set are shown dotted.
    The newly inserted vertices appear as squares.}
    \label{fig:HPextended}
  \end{minipage}
\end{figure}

Let $G=(V,E)$ be an embedded  planar graph, let $E_c$ be an
HP-completion set of $G$ and let $\Gamma(G^\prime)$ of $G^\prime =
(V, E^\prime)$ be a drawing with $c$ crossings that preserves  $G$.
The graph $G_c$ induced from drawing $\Gamma(G^\prime)$ by inserting
a new vertex at each edge crossing and by splitting the edges
involved in the edge-crossing is referred to as the
\emph{HP-extended graph of $G$ w.r.t. $\Gamma(G^\prime)$}. (See
Figure~\ref{fig:HPextended})

In this paper, we present a linear time algorithm for solving the
Acyclic-HPCCM problem for outerplanar triangulated $st$-digrpahs.
Let $G=(V,E)$ be a  digraph. A vertex $s \in V$ with in-degree equal
to zero (0)  is called a \emph{source}, while, a vertex $t \in V$
with outdegree equal to zero is called a  \emph{sink}. A
\emph{planar $st$-digraph} is an embedded  planar acyclic digraph
with exactly one source $s$ (i.e., a vertex with in-degree equal to
0) and exactly one sink $t$ (i.e., a vertex with out-degree equal to
0) both of which appear on the boundary of the external face. It is
known that a planar $st$-digraph admits a planar upward drawing
\cite{Kelly87,DiBattistaT88}. In the rest of the paper, all
$st$-digraphs will be drawn upward. A planar graph $G$ is
\emph{outerplanar} if there exist a drawing of $G$ such that all of
$G$'s vertices appear on the boundary of the same face (which is
usually drawn as the external face). A \emph{triangulated
outerplanar} graph is an outerplanar graph with triangulated
interior, i.e., all interior faces consist of 3 vertices and 3
edges.

\subsection{Related Work}

For acyclic digraphs, the  Acyclic-HPC problem has been studied in
the literature in the context of partially ordered sets (posets)
under the terms  \emph{Linear extensions} and \emph{Jump Number}.
Each acyclic-digraph $G$ can be treated as a poset $P$. A linear
extension of $P$ is a total ordering $L =\{x_1 \ldots x_n\}$ of the
elements of $P$ such that $x_i < x_j$ in $L$ whenever $x_i < x_j$ in
$P$. We denote by $L(P)$ the set of all linear extensions of $P$. A
consecutive pair $(x_i, x_{i+1})$ is called a \emph{\emph{jump} in
$L$} if $x_i$ is not comparable to $x_{i+1}$ in $P$. Denote the
number of jumps of $L$  by $s(P, L)$. Then, the \emph{jump number}
of $P$, $s(P)$, is defined as $s(P) = \min \{s(P, L): L \in L(P)\}$.
Call  a linear extension $L$ in $L(P)$ optimal if $s(P, L) = s(P)$.
The \emph{jump number problem} is to find $s(P)$ and to construct an
optimal linear extension of $P$.

From the above definitions, it follows that an optimal linear
extension of a poset $P$ (or its corresponding  acyclic digraph
$G$), is identical to an acyclic HP-completion set $E_c$ of minimum
size for $G$, and its jump number is equal to the size of $E_c$.
This problem has been widely studied, in part due to its
applications to scheduling. It has been shown to be NP-hard even for
bipartite ordered sets~\cite{Pulleyblank81} and the class of
interval orders~\cite{Mitas91}. Up to our knowledge, its
computational classification is still open for lattices.
Nevertheless, polynomial time algorithms are known for several
classes of ordered sets. For instance, efficient algorithms are
known for series-parallel orders~\cite{CogisH79}, N-free
orders~\cite{Rival83}, cycle-free orders~\cite{DuffusRW82}, orders
of width two~\cite{CheinH84}, orders of bounded
width~\cite{ColbournP85}, bipartite orders of dimension
two~\cite{SteinerS87} and K-free orders~\cite{ShararyZ91}.
Brightwell and  Winkler~\cite{BrightwellW91} showed that counting
the number of linear extensions is $\boldmath{\sharp}$P-complete. An
algorithm that generates all of the linear extensions of a poset in
a constant amortized time, that is in time $\mathcal{O}(|L(P)|)$,
was presented by Pruesse and Ruskey~\cite{PruesseR94}.

With respect to related work on book embeddings,
Yannakakis\cite{Yannakakis89} has shown that planar graphs have a
book embedding on a 4-page book and that there exist planar graphs
that require 4 pages for their book embedding. Thus, book embedding
for planar graphs are, in general, three-dimensional structures. If
we are interested only on two-dimensional structures we have to
concentrate on  2-page book embeddings and to allow spine crossings.
In the literature, the book embeddings where spine crossings are
allowed are referred to as \emph{topological book
embeddings}~\cite{EnomotoMO99}. It is known that every planar graph
admits a 2-page topological book embedding with only one spine
crossing per edge~\cite{DiGiacomoDLW05}.

For acyclic digraphs and posets, \emph{upward book embeddings} have
been also studied in the
literature~\cite{AlzohairiR96,HeathP97,HeathP99,HeathPT99,NowakowskiP89}.
An upward book embedding can be considered to be a book embedding in
which the spine is vertical and all edges are drawn monotonically
increasing in the upward direction.  The minimum number of pages
required by an upward book embedding of a planar acyclic digraph is
unbounded~\cite{HeathP97}, while, the minimum number of pages
required by an upward planar digraph is not
known~\cite{AlzohairiR96,HeathP97,NowakowskiP89}. Giordano et
al.~\cite{GiordanoLMS07} studied \emph{upward topological book
embeddings} of embedded upward planar digraphs, i.e., topological
2-page book embedding where all edges are drawn monotonically
increasing in the upward direction. They have showed how to
construct in linear time an upward topological book embedding for an
embedded triangulated planar $st$-digraph with at most one spine
crossing per edge. Given that (i) upward planar digraphs are exactly
the subgraphs of planar $st$-digraphs~\cite{DiBattistaT88,Kelly87}
and (ii) embedded upward planar digraphs they can be augmented to
become triangulated planar $st$-digraphs in linear
time~\cite{GiordanoLMS07}, it follows that any embedded upward
planar digraph has a topological book embedding with one spine
crossing per edge.

We emphasize that the presented bibliography is in no way
exhaustive. The topics of \emph{hamiltonian paths}, \emph{linear
orderings} and \emph{book embeddings} have been studied for a long
time and an extensive body of literature has been accumulated.


\section{Triangulated Hamiltonian $st$-Graphs}

In this section, we develop the necessary and sufficient condition
for a triangulated $st$-digraphs to be hamiltonian. The provided
characterization will be later on used in the development of
crossing-optimal HP-completion sets for outerplanar triangulated
$st$-digraphs.

It is well known\cite{TamassiaT86} that for every vertex $v$ of an
$st$-digraph, its incoming (outgoing) incident edges appear
consecutively around $v$. We denote by $Left(v)$ (resp. $Right(v)$)
the face to the left (resp. right) of the leftmost (resp. rightmost)
incoming and outgoing edges incident to $v$. The following lemma is
a direct consequence from a lemma by Tamassia and
Preparata~\cite[Lemma 7]{TamassiaP90}.

\begin{lemma}
\label{lem:TamassiaPreparata} Let $u$ and $v$ be two vertices of an
$st$-planar digraph such that there is no directed path between them
in either direction. Then, in the dual $G^*$ of $G$ there is either
a path from $Right(u)$ to $Left(v)$ or a path from $Right(v)$ to
$Left(u)$. \qed
\end{lemma}

The following lemma demonstrates a property of $st$-digraphs.

\begin{lemma}
\label{lemma:mutualDisconnected} Let $G$ be an st-digraph that does
not have a hamiltonian path. Then, there exist two vertices in $G$
that are not connected by a directed path in either direction.
\end{lemma}

\begin{proof}

Let $P$ be a  longest path from $s$ to $t$ and let $a$ be a vertex
that does not belong in $P$. Since $G$ does not have a hamiltonian
path, such a vertex always exists.  Let $s^\prime$ be the last
vertex in $P$ such that there exists a path
$P_{s^\prime\rightsquigarrow a}$  from $s^\prime$ to $a$ with no
vertices in $P$. Similarly, define $t^\prime$ to be the first vertex
in $P$ such that there exists a path $P_{ a\rightsquigarrow
t^\prime}$ from $a$ to $t^\prime$ with no vertices in $P$. Since $G$
is acyclic, $s^\prime$ appears before $t^\prime$ in $P$ (see
Figure~\ref{fig:mutualDisconnected}). Note that $s^\prime$ (resp.
$t^\prime$) might be vertex $s$ (resp. $t$). From the construction
of $s^\prime$ and $t^\prime$ it follows that any vertex $b$,
distinct from $s^\prime$ and $t^\prime$, that is located on path $P$
between vertices  $s^\prime$ and $t^\prime$, is not connected with
vertex $a$ in either direction. Thus, vertices $a$ and $b$ satisfy
the property of the lemma.

\begin{figure}[htb]
    \centering
    \includegraphics[width=0.2\textwidth]{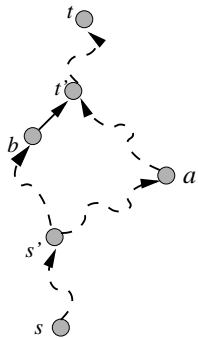}
    \caption{Subgraph used in the proof of
    Lemma~\ref{lemma:mutualDisconnected}. Vertices $a$ and $b$ are
    not connected by a path in either direction.}
    \label{fig:mutualDisconnected}
\end{figure}

Note that such a vertex $b$ always exists. If this was not the case,
then  path $P$ would contain edge $(s^\prime,t^\prime)$.  Then, path
$P$ could be extended by replacing  $(s^\prime, t^\prime)$ by path
$P_{s^\prime\rightsquigarrow a}$ followed by path
$P_{s^\prime\rightsquigarrow a}$. This would lead to new path
$P^\prime$ from $s$ to $t$ that is longer than $P$, a contradiction
since $P$ was assumed to be of maximum length. \qed
\end{proof}

\begin{figure}[htb]
    \begin{minipage}{0.32\textwidth}
    \centering
    \includegraphics[width=0.8\textwidth]{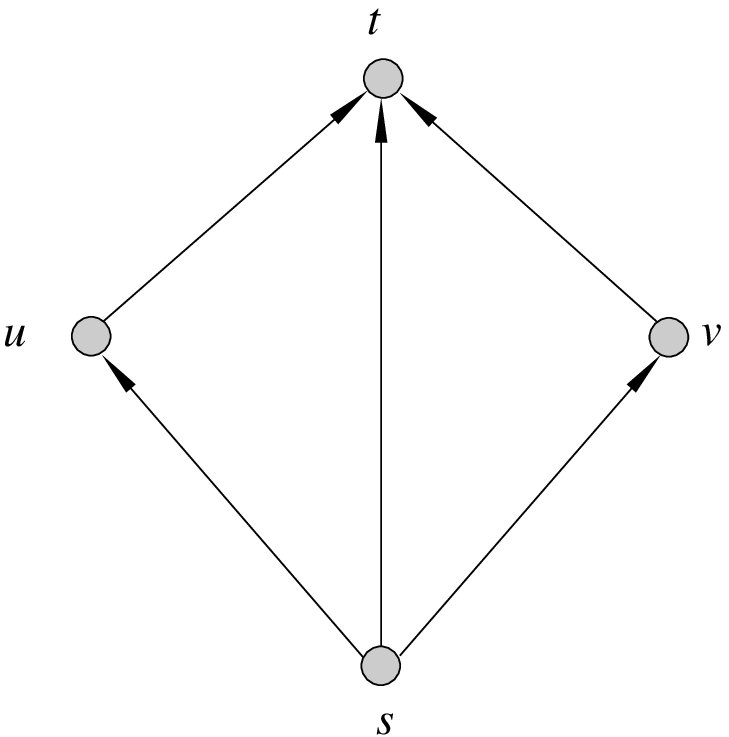}
    \caption{The rhombus embedded digraph.}
    \label{fig:rhombus}
  \end{minipage}
\hfill
    \begin{minipage}{0.32\textwidth}
    \centering
    \includegraphics[width=.75\textwidth]{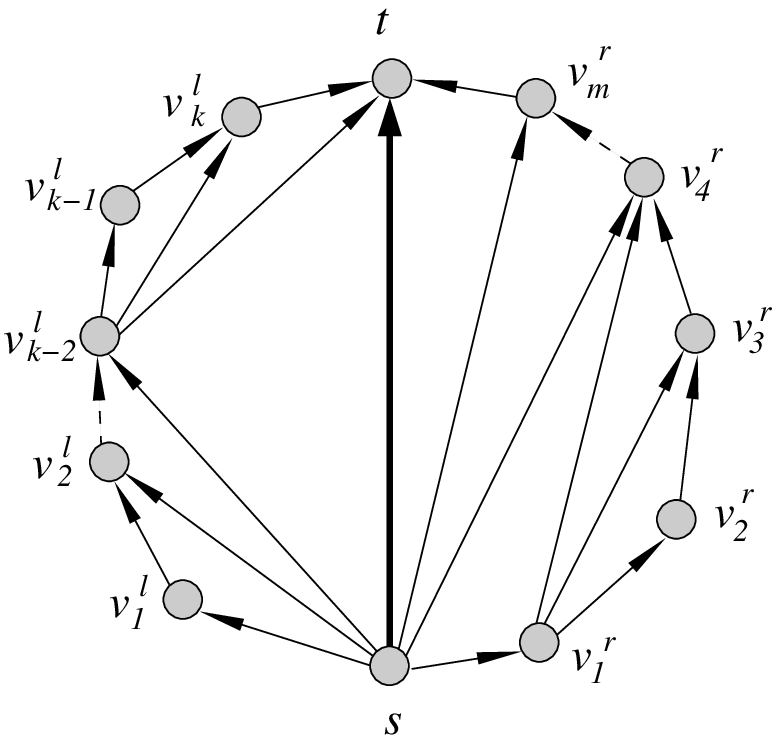}
    \caption{An $st$-polygon.}
    \label{fig:st-polygon}
  \end{minipage}
\hfill
    \begin{minipage}{0.32\textwidth}
    \centering
    \includegraphics[width=0.6\textwidth]{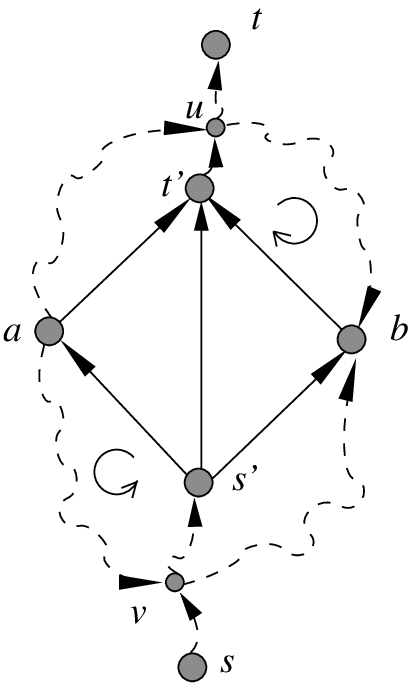}
    \caption{The $st$-digraph used in the proof of Theorem~\ref{lemma:mutualDisconnected}.}
    \label{fig:rhombusProof}
  \end{minipage}
\end{figure}

 The embedded
digraph in Figure~\ref{fig:rhombus} is called a \emph{rhombus}. The
central edge $(s,t)$ of a rhombus is referred to as the \emph{median
of the rhombus} and is always drawn in the interior of its drawing.

 An \emph{$st$-polygon} is a triangulated
outerplanar $st$-digraph that always contains edge $(s,t)$
connecting its source to its sink. Edge $(s,t)$ is referred to as
the \emph{median of the $st$-polygon} and it always lies in the
interior of the its drawing. As a consequence, an $st$-polygon must
have at least 4 vertices. Figure~\ref{fig:st-polygon} shows an
$st$-polygon. An $st$-polygon (that is a subgraph of some embedded
planar digraph) which cannot be extended by the addition of more
vertices to its external boundary is called a \emph{maximal
$st$-polygon}.

\begin{lemma}
\label{lemma:oneRombus} An $st$-polygon contains exactly one
rhombus.
\end{lemma}
\begin{proof}
By   definition, an $st$-digraph is a triangulated outerplanar graph
with $s$ and $t$ as its source and sink, respectively. It follows,
that the median of the $st$-polygon is also a median of a rhombus
formed by the two faces to the left and the right of edge $(s,t)$.
Assume now that there exists another rhombus. Then, due to the fact
that the $st$-polygon is outerplanar, all of its vertices would lie
on one side of the $(s,t)$-polygon. However, this is not possible
without violating the acyclicity of the $st$-polygon.
 \qed
\end{proof}

 The following theorem provides a characterization of
triangulated $st$-digraphs that have a hamiltonian path.

\begin{theorem}
\label{thm:rhombus}
Let $G$ be a triangulated $st$-digraph. $G$ has a hamiltonian path
if and only if $G$ does not contain any rhombus as a subgraph.
\end{theorem}

\begin{proof}

$(\Rightarrow)$~~~  We assume that $G$ has a hamiltonian path and we
will show that it contains no rhombus as a subgraph. For the sake of
contradiction,  assume that $G$ contains a rhombus composed from
vertices $s^\prime, ~t^\prime, ~a~ \mbox{and}~ b$ as a subgraph (see
Figure~\ref{fig:rhombusProof}). Then, vertices $a$ and $b$  of the
rhombus are not connected by a directed path in either direction. To
see this, assume wlog that there was a path connecting $a$ to $b$.
Then, this path has to intersect either the path from $t^\prime$ to
$t$ at a vertex $u$ or the path from $s$  to $s^\prime$ at a vertex
$v$. In either case, there must exist a cycle in $G$, contradicting
the fact that $G$ is acyclic. So, we have shown that vertices $a$
and $b$ of the rhombus are not connected  by a directed path in
either direction, and thus there cannot exist any hamiltonian path
in $G$, a clear contradiction.

$(\Leftarrow)$ We assume that $G$ contains no rhombus as a subgraph
and we will prove that $G$ has a hamiltonian path. For the sake of
contradiction, assume that $G$ does not have a hamiltonian path.
Then, from Lemma~\ref{lemma:mutualDisconnected}, if follows that
there exist two vertices $u$ and $v$ of $G$ that are not connected
by a directed path in either direction. From
Lemma~\ref{lem:TamassiaPreparata}, it then follows that there exists
in the dual $G^*$ of $G$ a directed path from either $Right(u)$ to
$Left(v)$, or from $Right(v)$ to $Left(u)$. Wlog, assume that the
path in the dual $G^*$ is from $Right(u)$ to $Left(v)$  (see
Figure~\ref{fig:rhombusProofBack}.a) and let $f_0, ~f_1, ~\ldots,
~f_k$ be the faces the path passes through, where $f_0=Right(u)$ and
$f_k=Left(v)$. We denote the path from $Right(u)$ to $Left(v)$ by
$P_{u,v}$.

\begin{figure}[tb]
    \begin{minipage}{\textwidth}
    \centering
    \includegraphics[width=1\textwidth]{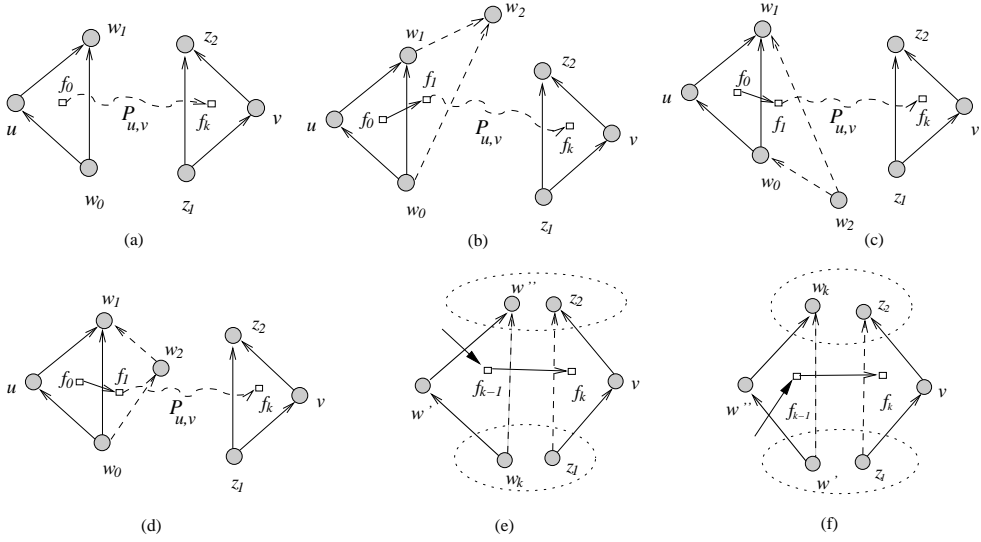}
    \caption{The different cases occurring in  the construction of path $P_{u,v}$ as described in
    the proof of Theorem~\ref{thm:rhombus}.}
    \label{fig:rhombusProofBack}
  \end{minipage}
\end{figure}

Note that path $P_{u,v}$ can exit face
$f_0=\vartriangle\hspace*{-5pt}(u,w_0,w_1)$ only through edge
$(w_0,w_1)$ (see Figure~\ref{fig:rhombusProofBack}.a). The path will
enter a new face and, in the rest of the proof, we will construct
the sequence of faces it goes through.

The next face $f_1$ of the path,  consists of edge $(w_0,w_1)$ which
is connected to a vertex $w_2$. For face
$f_1=\vartriangle\hspace*{-5pt}(w_0,w_1, w_2)$ there are 3 possible
orientations (which do not violate acyclicity) for the direction of
the two edges that connect $w_2$ with $w_0$ and $w_1)$:\\
\textbf{Case~1}: Vertex $w_2 $ has incoming edges from both $w_0$
and $w_1$ (see Figure~\ref{fig:rhombusProofBack}.b). Observe that
path $P_{u,v}$ can continue from $f_1$ to $f_2$ only by crossing
edge $(w_0, w_2)$. This is due to the fact that, in the dual $G^*$,
the only outgoing edge of $f_1$ corresponds to the dual edge that
crosses edge $(w_0, w_2)$ of $G$. \\
\textbf{Case~2}: Vertex $w_2 $ has outgoing edges to both $w_0$ and
$w_1$ (see Figure~\ref{fig:rhombusProofBack}.c). Observe that path
$P_{u,v}$ can continue from $f_1$ to $f_2$ only by crossing edge
$(w_2, w_0)$. This is due to the fact that, in the dual $G^*$, the
only outgoing edge of $f_1$ corresponds to the dual edge that
crosses edge $(w_2, w_0)$ of $G$. \\
\textbf{Case~3}: Vertex $w_2 $ is connected to   $w_0$ and $w_1$ by
an incoming and an outgoing edge, respectively (see
Figure~\ref{fig:rhombusProofBack}.d). Note that in this case, $f_0$
and $f_1$ form a rhombus. Thus, this case cannot occurs, since we
assumed that $G$ has no rhombus as a subgraph.

A common characteristic of either of the first two case that allow
to further continue the identification of the faces path  $P_{u,v}$
goes through is that there is \emph{a single} edge that can be used
to exit face $f_1$. Thus, we can continue identifying the faces path
$P_{u,v}$  passes through, building in such a way sequence $f_0,
~f_1, ~\ldots, ~f_{k-1}$.

At the end, path $P_{u,v}$ has to leave face $f_{k-1}$ by either
edge $(w_k, w^{\prime\prime})$ (see
Figure~\ref{fig:rhombusProofBack}.e) or by edge $(w^{\prime}, w_k)$
(see Figure~\ref{fig:rhombusProofBack}.f). In either of these two
cases, the outgoing boundary edge of face
$f_{k-1}=\vartriangle\hspace*{-5pt}(w^{\prime},w^{\prime\prime},
w_k)$ has to be identified with the single incoming edge of face
$f_{k-1}=\vartriangle\hspace*{-5pt}(z_1,z_2, v)$. Thus, the last two
faces on the path $P_{u,v}$ will form a rhombus (see
Figures~\ref{fig:rhombusProofBack}.e and
~\ref{fig:rhombusProofBack}.f). Thus, in either case, in order to
complete the path, graph $G$ must contain a rhombus as a subgraph.
This is a clear contradiction since we assumed that $G$ does not
contain any rhombus as a subgraph.\qed
\end{proof}

\section{Optimal Acyclic Hamiltonian Path Completion for Outerplanar Triangulated st-digraphs}
Let $G=(V^l \cup V^r \cup \{s,t\}, E)~$ be an outerplanar
triangulated $st$-digraph, where $s$ is its source, $t$ is its sink
and  the vertices in $V_l$ (resp. $V_r$) are located on the left
(resp. right) part of the boundary of the external face. Let $V^l =
\{ v^l_1,~\ldots, v^l_k\}$ and $V^r = \{  v^r_1,~\ldots, v^r_m\}$,
where the subscripts indicate the order in which the vertices appear
on the left (right) part of the external boundary. By convention,
source and the sink are considered to lie on both the left and the
right sides of the external boundary. For brevity, we refer to an
\underline{o}uterplanar \underline{t}riangulated $st$-digraph, as
\emph{OT-$st$-digraph}.

In this section we present an algorithm that computes a
crossing-optimal acyclic HP-completion set  for an OT-$st$-digraph
when only one crossing per edge  is permitted.

\subsection{$st$-polygon decomposition of an OT-$st$-digraph}

Observe that an $st$-polygon fully covers a strip of the
OT-$st$-digraph that is defined by two edges (one adjacent to its
source and one to its sink), each having its endpoints at different
sides of $G$. We refer to these two edges as the \emph{limiting
edges} of the $st$-polygon.

\begin{lemma}
\label{lem:medianComputation} Given an OT-$st$-digraph $G=(V^l \cup
V^r \cup \{s,t\}, E)~$ and an edge $e=(s^\prime, t^\prime) \in E$,
we can decide in $O(1)$ time whether $e$ is a median edge of some
$st$-polygon. Moreover, the two vertices (in addition to $s^\prime$
and $t^\prime$) that define
 a maximal $st$-polygon that has edge $e$ as its median can be also computed in $O(1)$ time.
\end{lemma}

\begin{proof}

We refer to a edge that has both of each end-vertices on the same
side of $G$ as a \emph{one-sided} edge. All remaining edges are
referred to as \emph{two-sided} edges. The edges exiting the source
and the edges entering the sink are treated as one-sided edges.

\begin{figure}[htb]
    \begin{minipage}{0.45\textwidth}
    \centering
    \includegraphics[width=0.5\textwidth]{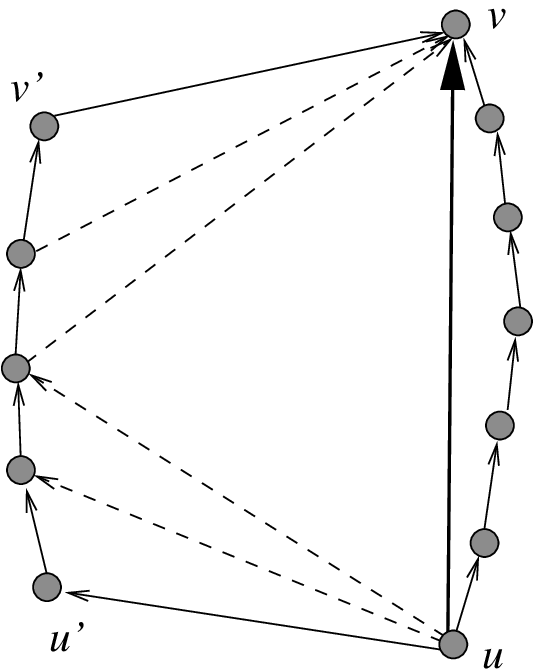}
    \caption{The solid edges are the edges that bounds the $st$-polygon and its median.}
    \label{fig:detect1}
  \end{minipage}
  \hfill
  \begin{minipage}{0.45\textwidth}
    \centering
    \includegraphics[width=0.5\textwidth]{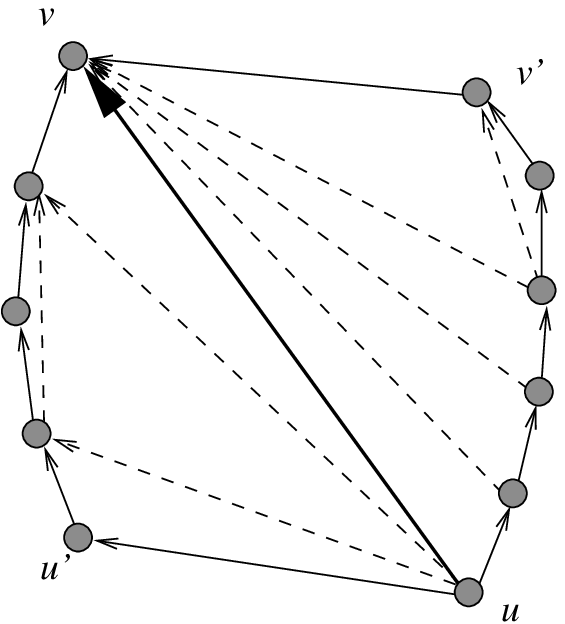}
    \caption{The solid edges are the edges that bounds the $st$-polygon and its median.}
    \label{fig:detect2}
  \end{minipage}
\end{figure}

Observe that a one-sided edge $(u,v)$ is a median of an $st$-polygon
if the following hold (see Figure~\ref{fig:detect1}):
\vspace*{-0.4cm}
\begin{enumerate}[(i)]
\item $u$ and $v$ are not successive vertices of the side of $G$.
\item $u$ has a two-sided outgoing edge.
\item $v$ has a two-sided incoming edge.
\end{enumerate}

Similarly, observe that a two-sided edge $(u,v)$ with $u \in V^R$
(resp.  $u \in V^L$) is a median of an $st$-polygon if the following
hold (see Figure~\ref{fig:detect2}): \vspace*{-0.4cm}
\begin{enumerate}[(i)]
\item $u$ has a two-sided outgoing edge that is clock-wise before (resp. after)
$(u,v)$ .
\item $v$ has a two-sided incoming edge that is clock-wise before $(u,v)$.
\end{enumerate}

All of the above conditions can be trivially tested in $O(1)$ time.
We can preprocess graph $G$ in linear time so that for each of its
vertices  we know its first and last (in clock-wise order) in-coming
and out-going edges, Then, the two remaining vertices that define
the maximal $st$-polygon having $(u,v)$ as its median can be found
in $O(1)$ time and it can be reported in time proportional to its
size. \qed
\end{proof}

\begin{lemma}
\label{lem:areaDisjointPolygons}
 The $st$-polygons of an
OT-$st$-digraph $G$ are mutually area-disjoint.
\end{lemma}
\begin{proof}
From Lemma~\ref{lemma:oneRombus} it follows that one $st$-polygon
cannot fully contain another. Thus, we have to consider only partial
area overlap. For the sake of contradiction assume that there are
two $st$-polygons, say $R_1$ and $R_2$  that overlap. Then they have
to share at least a face. Thus, a partial area overlap would force
one endpoint of a limiting edge of $R_1$ to be located inside $R_2$
(or vice versa), and more precisely on the boundary nodes between
the two limiting edges of $R_2$. As a consequence, we would have
either a crossing between the limiting edges of the two
$st$-polygons or between a median of one $st$-polygon and a limiting
edge of another.\qed
\end{proof}

\begin{figure}[htb]
    \centering
    \includegraphics[width=1\textwidth]{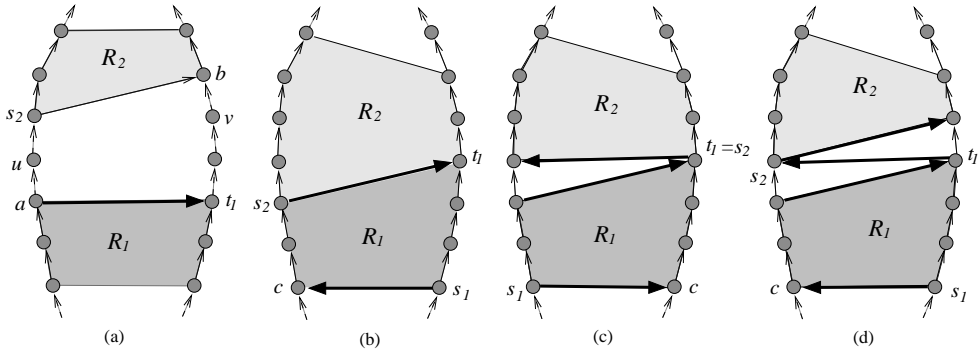}
    \caption{Configurations of adjacent $st$-polygons of an OT-$st$-digraph.}
    \label{fig:succesivePolygons}
\end{figure}

Denote by $\mathcal{R}(G)$ the set of all maximal $st$-polygons of
an OT-$st$-digraph $G$.  Observe than not every vertex of $G$
belongs to one of its maximal $st$-polygons. We refer to  the
vertices of $G$ that are not part of any maximal $st$-polygon as
\emph{free vertices} and we denote them by $\mathcal{F}(G)$.
 Also observe that
an ordering can be imposed on the maximal $st$-polygons of an
OT-$st$-digraph $G$ based on the ordering of the area disjoint
strips occupied by each $st$-polygon. The vertices which do not
belong to an $st$-polygon are located between the strips occupied by
consecutive $st$-polygons (see
Figure~\ref{fig:succesivePolygons}.a).

\begin{lemma}
\label{lem:pahtProperties}
 Assume an OT-$st$-digraph $G$. Let
$R_1$ and $R_2$ be two of $G$'s consecutive maximal $st$-polygons
and let $V_f \subset \mathcal{F}(G)$ be the set of free vertices
lying between $R_1$ and $R_2$. Then, the following statements are
satisfied: \vspace*{-.5cm}
\begin{enumerate}[(i)]
\item For any pair of vertices $u,~v\in V_f$ there is either a path
from $u$ to $v$ or from $v$ to $u$.
\item For any vertex $v\in V_f$ there is a path from the sink of $R_1$
to $v$ and from $v$ to the source of $R_2$.
\item If $V_f=\emptyset$, then there is a path from source of $R_1$
to the source of $R_2$.
\end{enumerate}
\end{lemma}

\begin{proof}
$~$\\\vspace*{-1.0cm}
\begin{enumerate}[(i)]
\item
For the sake of contradiction, assume that $u$ and $v$ are not
connected by a path in either direction. Observe that the subgraph
$G_f$ of $G$ induced by the vertices of $V_f$ together with  the
vertices of the   upper limiting edge of $R_1$ and the lower
limiting edge of $R_2$ is a triangulated $st$-digraph. By
Theorem~\ref{thm:rhombus}, if follows that $G_f$ contains a rhombus
(and thus an $st$-polygon). This is a contradiction since we assumed
that $R_1$ and $R_2$ are consecutive.

\item Denote by $t_1$ the sink of $R_1$. Assume that $v$ and $t_1$ lie on different sides of $G$, otherwise there is
always a path form $t_1$ to $v$ (see
Figure~\ref{fig:succesivePolygons}.a). Observe that edge $(v, t_1)$
does not exist in $G$; if it existed, it would be the upper limiting
edge of $R_1$, not $(a, t_1)$. If we also assume the edge $(t_1, v)$
does not exist, then $v$ and $t_1$ are not connected by a path in
either direction. Then the argument continues as in case (i),
leading to a contradiction. The claim that there is a path from $v$
to the source $s_2$ of $R_2$ is similar.

\item Note that there are 3 configuration  in which no free vertex exists between two consecutive
$st$-polygons (see Figures~\ref{fig:succesivePolygons}.b-d). Denote
by $s_1$ and $s_2$ the sources of $R_1$ and $R_2$, respectively. If
$s_1$ and $s_2$ lie on the same side of $G$ then the claim is
obviously true since $G$ is an OT-$st$-digraph. If they belong to
opposite sides of $G$, observe that the lower limiting edge
$(s_1,c)$ of $R_1$ leads to the side of $G$ which contains $s_2$.
Since there is a path from $c$ to $s_2$, it follows that there is a
path from $s_1$ to $s_2$.\qed
\end{enumerate}
\end{proof}

We refer to the source vertex $s_i$ of each maximal $st$-polygon
$R_i \in \mathcal{R}(G),~ 1 \leq i \leq |\mathcal{R}(G)|$ as the
\emph{representative} of $R_i$ and we denote it by $r(R_i)$. We also
define the representative of a free vertex $v\in \mathcal{F}(G)$ to
be $v$ itself, i.e. $r(v)=v$. For any two distinct elements $x, ~y~
\in \mathcal{R}(G) \cup \mathcal{F}(G)$, we define the  relation
$\angle_p$ as follows: \emph{$x \angle_p y$ iff there exists a path
from}  $r(x)$ \emph{to} $r(y)$.

\begin{lemma}
\label{lem:totalOrder}
 Let $G$ be an $n$ node OT-$st$-digraph. Then,
relation $\angle_p$ defines a total order on the elements
$\mathcal{R}(G) \cup \mathcal{F}(G)$. Moreover, this total order can
be computed in $O(n)$ time.
\end{lemma}

\begin{proof}
The fact that $\angle_p$ is a total order on $\mathcal{R}(G) \cup
\mathcal{F}(G)$
 follows from Lemma~\ref{lem:pahtProperties}. The  order of
 the element of $\mathcal{R}(G) \cup
\mathcal{F}(G)$ can be easily derived by the
 numbers assigned to the representatives of the elements (i.e., to
 vertices of $G$) by a topological sorting of the vertices of $G$.
 To complete the proof, recall that an $n$ node acyclic planar graph can be
 topologically sorted in $O(n)$ time.
\qed
\end{proof}

\begin{definition}
Given an OT-$st$-digraph $G$,  the \emph{$st$-polygon decomposition}
$\mathcal{D}(G)$ of $G$ is defined to be the total order of its
maximal $st$-polygons and its free vertices induced by relation
$\angle_p$.
\end{definition}

The following theorem follows directly from
Lemma~\ref{lem:medianComputation} and Lemma~\ref{lem:totalOrder}.

\begin{theorem}
\label{thm:STpolygonDecomposition} An $st$-polygon decomposition of
an $n$ node OT-$st$-digraph $G$ can be computed in $O(n)$ time.
\end{theorem}

\subsection{The Algorithm}

The following lemmata concern a crossing-optimal acyclic
HP-completion set for a single $st$-polygon and are essential for
the development of our   algorithm for OT-$st$-digraphs.

\eat{ Let $R$ be an $st$-polygon $R=(V^l \cup V^r \cup \{s,t\}, E)~$
where $V_l$ and $V_r$ are the vertices to its left and right border,
respectively.
 }

\begin{lemma}
\label{lem:NoMixedSidesBook} Assume an $st$-polygon $R=(V^l \cup V^r
\cup \{s,t\}, E)~$. In a crossing-optimal acyclic HP-completion set
of $R$ with at most 1 edge crossing per initial edge,  the vertices
of
  $V^l$ are visited before the vertices of $V^r$, or, vice
  versa.
\end{lemma}

\begin{proof}
In an $st$-polygon, due to the existence of the median $(s,t)$, no
edges exists between vertices of $V^l$ and $V^r$. Thus, the edges of
a Hamiltonian path that connect vertices at different sides all
belong in the HP-completion set and they all cross the median. Since
we insist on having at most one crossing per edge of $R$, the
HP-completion set must consist of a single edge. This implies that
all vertices of $V^l$ are visited before the vertices of $V^r$, or,
vice versa.\qed
\end{proof}

\begin{lemma}
\label{lem:spineCrossingSTpolygon} Assume an $n$ node  $st$-polygon
$R=(V^l \cup V^r \cup \{s,t\}, E)~$. A crossing-optimal  acyclic
HP-completion set for $R$ with at most 1 edge crossing per initial
edge can be computed in $O(n)$ time.
\end{lemma}

\begin{proof}
\begin{figure}[htb]
    \centering
    \includegraphics[width=.5\textwidth]{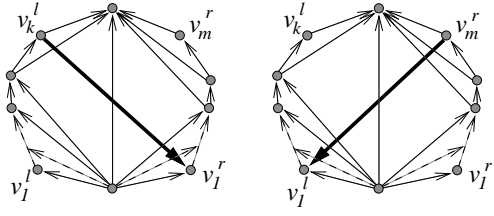}
    \caption{Edge crossings in crossing-optimal HP-completion of an $st$-polygon.}
    \label{fig:HPcompletionPolygon}
\end{figure}

Let $V^l = \{ v^l_1,~\ldots, v^l_k\}$ and $V^r = \{  v^r_1,~\ldots,
v^r_m\}$, where the subscripts indicate the order in which the
vertices appear on the left (right) boundary of $R$. Suppose that
$I:V \times V \rightarrow \{0,1\}$ is an indicator function such
that $I(u,v)=1 \iff (u,v) \in E$. Due to
Lemma~\ref{lem:NoMixedSidesBook}, the single edge of the
HP-Completion set is either edge $(v^l_k, v^r_1)$ or $(v^r_m,
v^l_1)$. Edge $(v^l_k, v^r_1)$ has to cross  all edges connecting
$t$ with vertices of $V_l \setminus \{v^l_k\} $, the median, and all
edges connecting $s$ with vertices of $V_r\setminus \{v^R_1\}$ (see
Figure~\ref{fig:HPcompletionPolygon}.a). Similarly, Edge $(v^r_m,
v^l_1)$ has to cross  all edges connecting $t$ with vertices of $V_r
\setminus \{v^r_m\}$, the median, and all  edges connecting $s$ with
vertices of $V_l\setminus \{v^l_1\}$(see
Figure~\ref{fig:HPcompletionPolygon}.b). Then, the edge involved in
the minimum number of crossings can be computed in $O(n)$ time and
the corresponding number of crossings is:
$$1 + \min\left\{  \sum_{i=1}^{k}{I(v^\ell_i,t)}+\sum_{i=1}^{m}{I(s,v^r_i)}~,
             ~\sum_{i=1}^{m}{I(v^r_i,t)}    +\sum_{i=1}^{k}{I(s,v^\ell_i)}\right\}$$
\qed
\end{proof}

Assume an OT-$st$-digraph $G$. We denote by $S(G)$ the hamiltonian
path on the HP-extended digraph of $G$ that results when a
crossing-optimal HP-completion set is added to $G$. Note that if we
are only given $S(G)$ we can infer the size of the HP-completion set
and the number of edge crossings. Denote by $c(G)$ the number of
edge crossings caused by the HP-completion set inferred by $S(G)$.
If we are restricted to Hamiltonian paths that enter the sink of G
from a vertex on the left (resp. right) side of $G$, then we denote
the corresponding size of HP-completion set as $c(G,L)$ (resp.
$c(G,R)$). Obviousy, $c(G)= \min \{ c(G,L),~c(G,R) \}$.

We use the operator $\oplus$ to indicate the concatenation of two
paths. By convention, the hamiltonian path of a single vertex is the
vertex itself.

Let $\mathcal{D}(G)= \{ o_1,~ \ldots, o_\lambda \}$ be the
$st$-polygon decomposition of $G$, where element $o_i,~ 1\leq i \leq
\lambda$ is either an $st$-polygon or a free vertex. Denote by
$G_i,~1\leq i \leq \lambda $ to be the graph induced by the vertices
of elements $o_1, \ldots, o_i$. Observe that graph $G_i$ is also an
OT-$st$-digraph. The same holds for the subgraph of $G$ that is
inferred by any number of consecutive elements in $\mathcal{D}(G)$.

\begin{lemma}
\label{lem:optimalSubsolution} Assume an OT-$st$-digraph $G$ and let
$\mathcal{D}(G)= \{ o_1,~ \ldots, o_\lambda \}$ be its $st$-polygon
decomposition. Consider any two consecutive elements $o_i$ and
$o_{i+1}$ of $\mathcal{D}(G)$  that   share at most one vertex.
Then, the following statements hold:\vspace*{-.5cm}
\begin{enumerate}[(i)]
\item $S(G_{i+1})=S(G_i) \oplus S(o_{i+1})$,  and
\item $c(G_{i+1})= c(G_i)$.
\end{enumerate}
\end{lemma}

\begin{figure}[htb]
    \centering
    \includegraphics[width=1\textwidth]{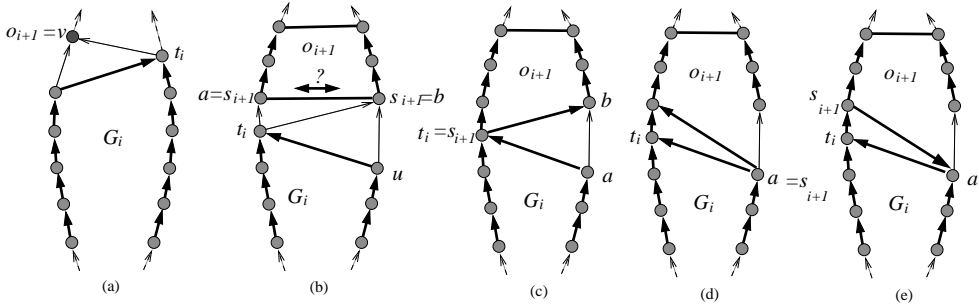}
    \caption{Configuration used in the proof of Lemma~\ref{lem:optimalSubsolution}.}
    \label{fig:optimalSubsolution}
\end{figure}

\begin{proof}
We proceed to prove first statement (i). There are three cases to
consider in which 2 consecutive elements of
$\mathcal{D}(G)$ share at most 1 vertex.\\
\textbf{Case-1}: Element $o_{i+1}=v$ is a free vertex (see
Figure~\ref{fig:optimalSubsolution}.a). By
Lemma~\ref{lem:pahtProperties}, if $o_i$ is either a free vertex or
an $st$-polygon, there is an edge connecting the sink of $o_i$ to
$v$. Also observe that if $v$ was not the last vertex of
$S(G_{i+1})$ then the crossing-optimal HP-completion set had to
include an edge from $v$ to some vertex of $G_i$. This is impossible
since it would create a cycle in the HP-extended digraph of
$S(G_{i+1})$.

\textbf{Case-2}: Element $o_{i+1}$ is an $st$-polygon that shares no
common vertex with $G_i$ (see
Figure~\ref{fig:optimalSubsolution}.b). Without loss of generality,
assume that the sink of $G_i$ is located on its left side. We first
observe that edge $(t_i, s_{i+1})$ exists in $G$. If $s_{i+1}$ is on
the left side of $G$, we are done. If $s_{i+1}$ is on the right side
of $G$, realize that edge $(u,a)$ cannot exist in $G$, since, if it
existed, the area between the two polygons would be an $st$-polygon
itself. Thus, that area can be only triangulated by edge $(t_i,
s_{i+1})$. Thus, as indicated in
Figure~\ref{fig:optimalSubsolution}.b, each of the end-vertices of
the lower limiting edge of $o_{i+1}$ can be its source. Since edge
$(t_i, s_{i+1})$ exists, the solution $S(o_{i+1})$ can be
concatenated to $S(G_i)$ and yield a valid hamiltonian path for
$G_{i+1}$. Now notice that in $S(G_{i+1})$ all vertices of $G_i$
have to be placed before the vertices of $o_{i+1}$. If this was not
the case, then the crossing-optimal HP-completion set had to include
an edge from a vertex $v$ of $o_{i+1}$ to some vertex $u$ of $G_i$.
This is impossible since it would create a cycle in the HP-extended
digraph of $S(G_{i+1})$.

\textbf{Case-3}: Element $o_{i+1}$ is an $st$-polygon that shares
one common vertex with $G_i$ (see
Figure~\ref{fig:optimalSubsolution}.c). Without loss of generality,
assume that the sink $t_i$ of $G_i$ is located on its left side.
Firstly, notice that the  vertex shared by $G_i$ and $o_{i+1}$ has
to be vertex $t_i$. To see that let $a$ be upper vertex at the right
side of $G_i$. Then, edge $(a,t_i)$ exists since $t_i$ is the sink
of $G_i$. For the sake of contradiction assume that $a$ was the
vertex shared between $G_i$ and $o_{i+1}$. If $a$ was also the
source of $o_{i+1}$ (see Figure~\ref{fig:optimalSubsolution}.d) then
$o_{i+1}$ wouldn't be maximal (edge $(a,t_i)$ should also belong to
$o_{i+1}$). If  $s_{i+1}$ was on the left side (see
Figure~\ref{fig:optimalSubsolution}.e), then a cycle would be formed
involving edges $(t_i),~(t_1, s_{i+1})$ and $(s_{i+1},a)$, which is
impossible since $G$ is acyclic. Thus, the vertex shared by $G_i$
and $o_{i+1}$ has to be vertex $t_i$. Secondly, observe that $t_i$
must coincide with vertex $s_{i+1}$ (see
Figure~\ref{fig:optimalSubsolution}.c). If $s_{i+1}$ coincided with
vertex $b$, then the $st$-polygon $o_i$ wouldn't be maximal since
edge $(b,t_i)$ should also belong to $o_{i}$. We conclude that $t_i$
coincides with $s_{i+1}$ and, thus, the solution $S(o_{i+1})$ can be
concatenated to $S(G_i)$ and yield a valid hamiltonian path for
$G_{i+1}$. To complete the proof for this case, we can show by
contradiction (on the acyclicity of $G$; as in Case-2)  that in
$S(G_{i+1})$ all vertices of $G_i$ have to be placed before the
vertices of $o_{i+1}$.

Now observe that statement (ii) is trivially true since, in all
three cases, the hamiltonian paths were extended by an edge of graph
$G$. Since $G$ is planar, no new crossings are created.\qed
\end{proof}

\begin{lemma}
\label{lem:dynProgSolution} Assume an  OT-$st$-digraph $G=(V^l \cup
V^r \cup \{s,t\}, E)~$ such that
 its  $st$-polygon decomposition
$\mathcal{D}(G)= \{ o_1,~ \ldots, o_\lambda \}$ does not contain any
free vertices and, moreover, all of $\mathcal{D}(G)$'s adjacent
elements share exactly two vertices.  Then, the following statements
hold:\vspace*{-.5cm}
\begin{enumerate}[(1)]
\item $~t_{i} \in V^l ~\Rightarrow~ c(G_{i+1},L)=\min \{
    c(G_{i},L)+c(o_{i+1},L)+1,~
    c(G_{i},R)+c(o_{i+1},L)
\} $
\item $~t_{i} \in V^l ~\Rightarrow~  c(G_{i+1},R)=\min \{
    c(G_{i},L)+c(o_{i+1},R),~
    c(G_{i},R)+c(o_{i+1},R)
\} $
\item $~t_{i} \in V^r ~\Rightarrow~  c(G_{i+1},L)=\min \{
    c(G_{i},L)+c(o_{i+1},L),~
    c(G_{i},R)+c(o_{i+1},L)
\} $
\item $~t_{i} \in V^r ~\Rightarrow~  c(G_{i+1},R)=\min \{
    c(G_{i},L)+c(o_{i+1},R),~
    c(G_{i},R)+c(o_{i+1},R)+1
\} $
\end{enumerate}
\end{lemma}

\begin{figure}[htb]
  \begin{minipage}{.47\textwidth}
    \centering
    \includegraphics[width=1\textwidth]{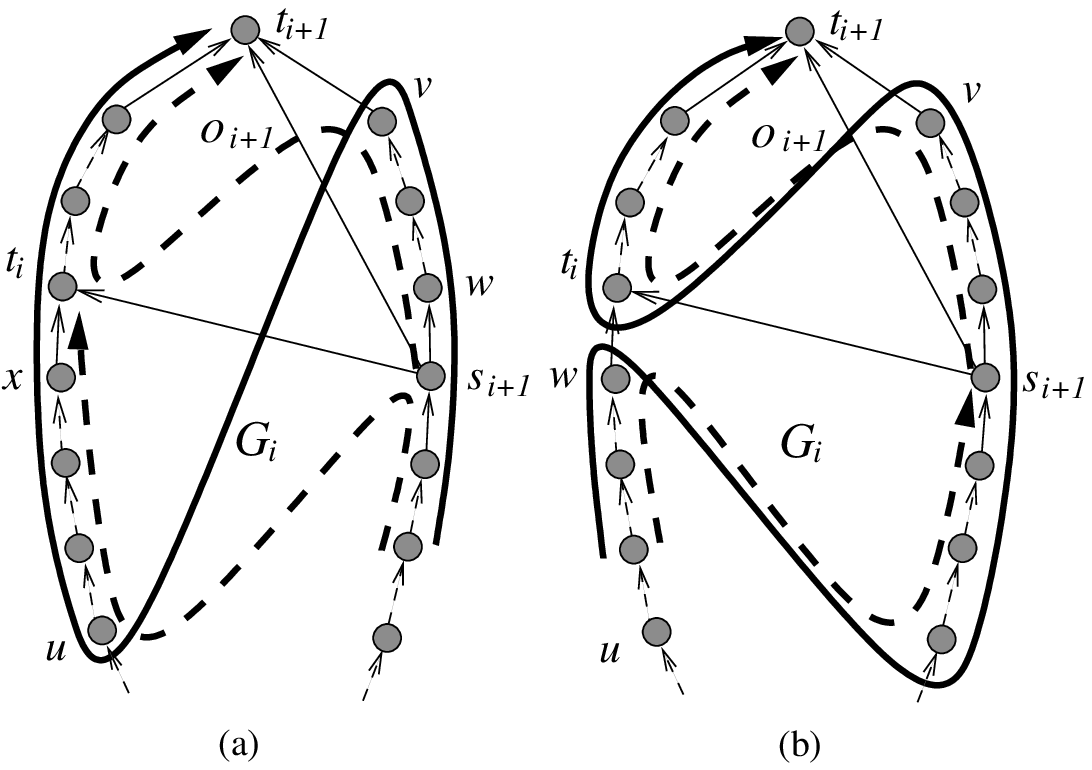}
    \caption{The hamiltonian paths for statement~(1) of Lemma~\ref{lem:dynProgSolution}.}
    \label{fig:dynProgSolutionLEFT-L}
  \end{minipage}
  \hfill
    \begin{minipage}{.47\textwidth}
    \centering
    \includegraphics[width=1\textwidth]{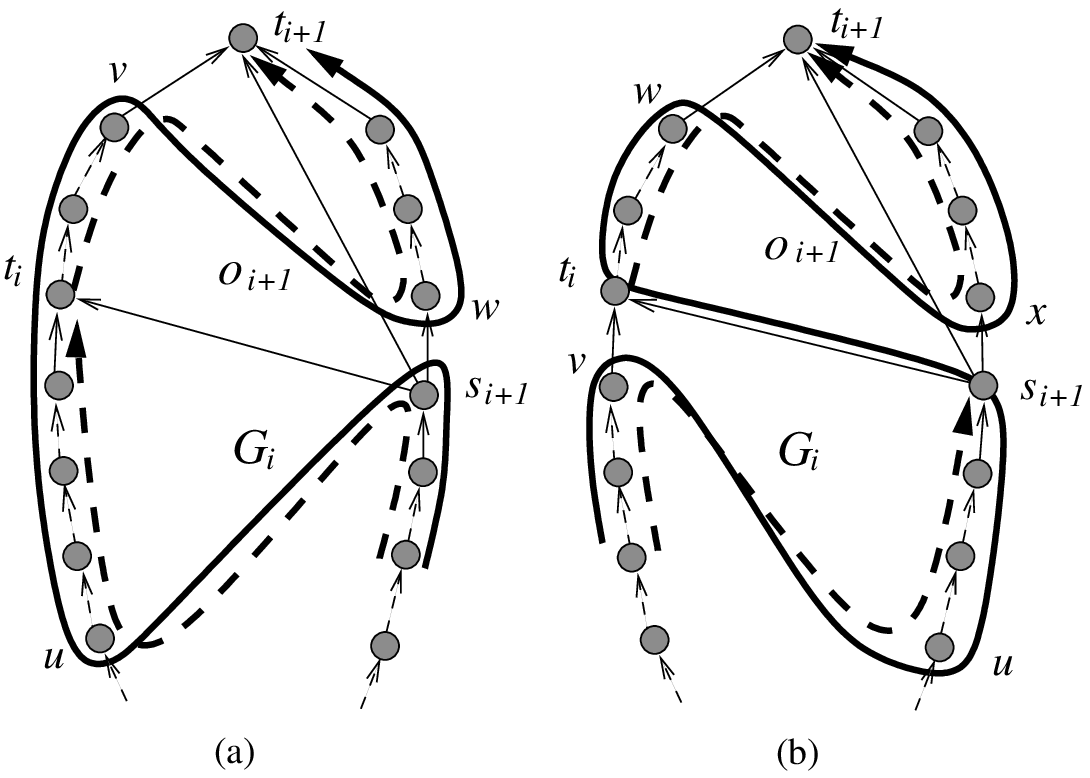}
    \caption{The hamiltonian paths for statement~(2) of Lemma~\ref{lem:dynProgSolution}.}
    \label{fig:dynProgSolutionLEFT-R}
  \end{minipage}
\end{figure}

\begin{proof}
We show how to build hamiltonian paths that infer HP-completions
sets of the specified size. Then, we prove that these HP-completion
sets are crossing-optimal. For each of the statements, there are two
cases to consider. The minimum number of crossings, is then
determined by taking the minimum over the two sub-cases.

 (1) $t_{i}
\in V^l \Rightarrow c(G_{i+1},L)=\min \{
    c(G_{i},L)+c(o_{i+1},L)+1,~
    c(G_{i},R)+c(o_{i+1},L)
\}$. 
\begin{quote}
\hspace*{-.4cm}\emph{Case 1a}. The hamiltonian path
  enters $t_{i}$ from a vertex on the left side of $G_i$.
 Figure~\ref{fig:dynProgSolutionLEFT-L}.a shows the resulting path.
 From the figure, it follows that $c(G_{i+1},L)=c(G_{i},L)+c(o_{i+1},L)+1$. To see that, just follow
 edge $(v,u)$ that becomes part of the completion set of $G_{i+1}$.
 Edge $(v,u)$ is involved in as many edge crossings as edge $(v,t_i)$
 (the only edge in the HP-completion set of $o_{i+1}$),
 plus as many edge crossings as edge $(s_{i+1},u)$
 (an edge in the HP-completion set of $G_{i}$), plus one (1) edge
 crossing of the lower
 limiting edge of $o_{i+1}$.\\
\hspace*{-.4cm}\emph{Case 1b}. The hamiltonian path  reaches
$t_{i}$
 from a vertex on the right side of $G_i$.
 Figure~\ref{fig:dynProgSolutionLEFT-L}.b shows the resulting path.
 From the figure, it follows that $c(G_{i+1},L)=
    c(G_{i},R)+c(o_{i+1},L)$.
 \end{quote}

 (2) $t_{i} \in V^l \Rightarrow  c(G_{i+1},R)=\min \{
    c(G_{i},L)+c(o_{i+1},R),~
    c(G_{i},R)+c(o_{i+1},R)
\} $. \vspace*{-.4cm}
 \begin{quote}
\hspace*{-.4cm}\emph{Case 2a}. The hamiltonian path
  enters $t_{i}$ from a vertex on the left side of $G_i$.
 Figure~\ref{fig:dynProgSolutionLEFT-R}.a shows the resulting path.
 From the figure, it follows that $c(G_{i+1},R)=c(G_{i},L)+c(o_{i+1},R)$.\\
\hspace*{-.4cm}\emph{Case 2b}. The hamiltonian path  reaches
$t_{i}$
 from a vertex on the right side of $G_i$.
 Figure~\ref{fig:dynProgSolutionLEFT-R}.b shows the resulting path.
 From the figure, it follows that $c(G_{i+1},R)=
    c(G_{i},R)+c(o_{i+1},R)$.
 \end{quote}
The proofs for statements (3) and (4) are symmetric to those of
statements (2) and (1), respectively. Figures
~\ref{fig:dynProgSolutionRIGHT-L}
and~\ref{fig:dynProgSolutionRIGHT-R} show how to construct the
corresponding hamiltonian paths in each of the cases.

\begin{figure}[htb]
  \begin{minipage}{.47\textwidth}
    \centering
    \includegraphics[width=1\textwidth]{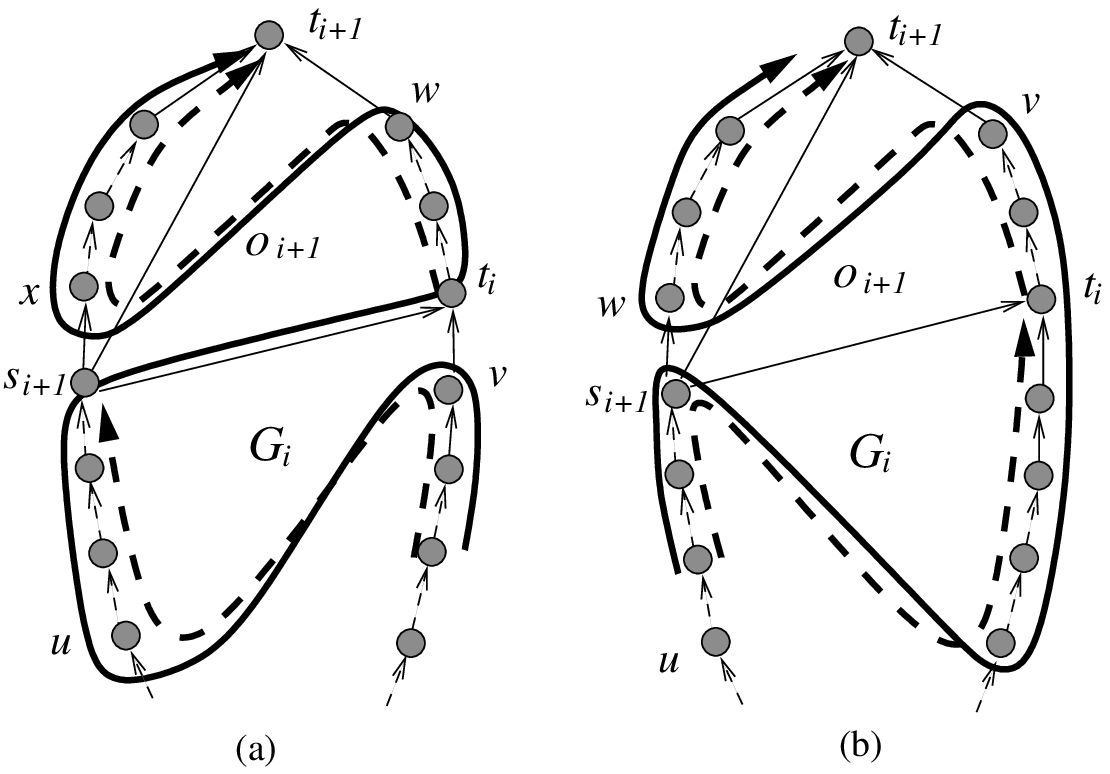}
    \caption{The hamiltonian paths for statement~(3) of Lemma~\ref{lem:dynProgSolution}.}
    \label{fig:dynProgSolutionRIGHT-L}
  \end{minipage}
  \hfill
    \begin{minipage}{.47\textwidth}
    \centering
    \includegraphics[width=1\textwidth]{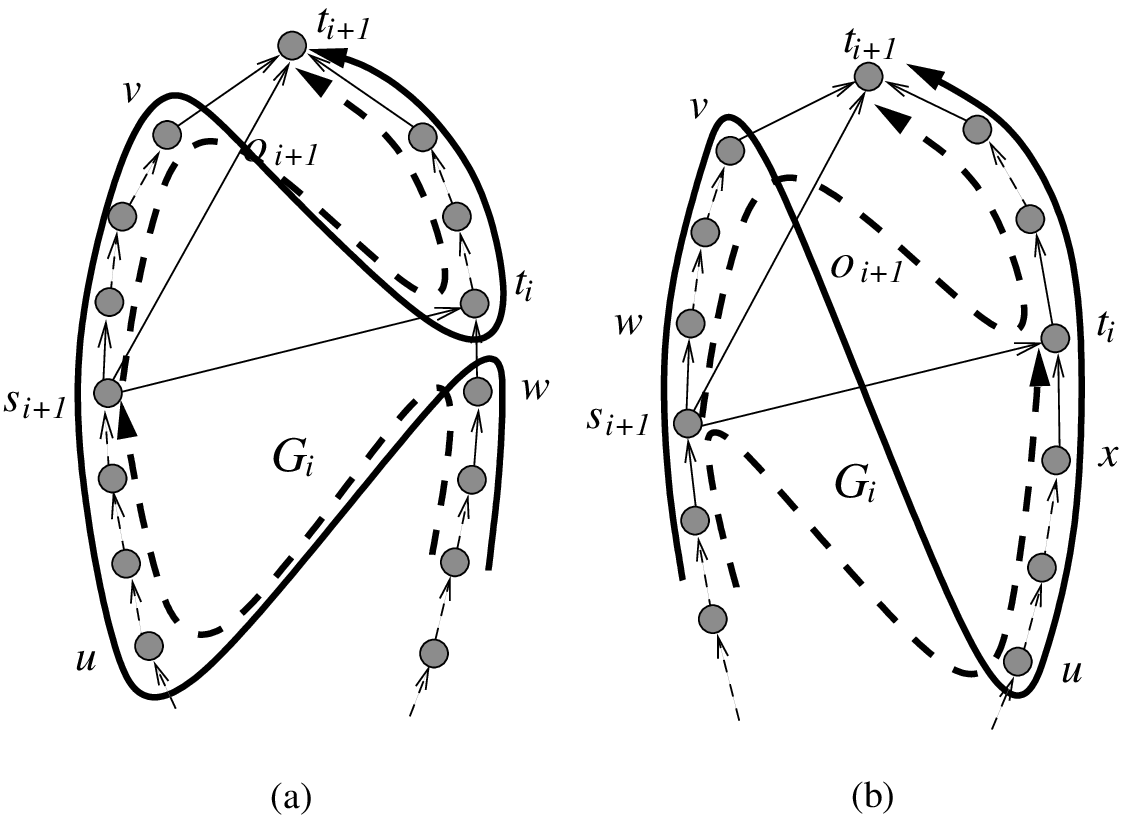}
    \caption{The hamiltonian paths for statement~(4) of Lemma~\ref{lem:dynProgSolution}.}
    \label{fig:dynProgSolutionRIGHT-R}
  \end{minipage}
\end{figure}

Denote by $l(G)$ (resp. $r(G)$) the sequence of vertices on the left
(resp. right) side of OT-$st$-digraph $G$. Note that $G$ can be an
$st$-polygon.

We now prove that, for each of the above cases, the  constructed
Hamiltonian paths  induce  crossing-optimal HP-completion sets of
the specified number of crossing. We show the proof just for Case~1a
(the most complicated, together with its symmetric Case~4b). The
proof for the other cases are similar and slightly simpler.

\begin{figure}[htb]
    \centering
    \includegraphics[width=1\textwidth]{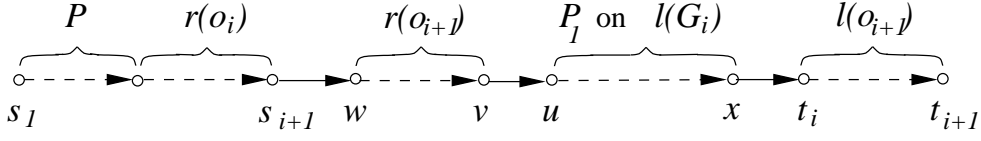}
    \caption{The structure of an acyclic-optimal hamiltonian path for Case~1a in the proof
    of Lemma~\ref{lem:dynProgSolution}.}
    \label{fig:dynProgSolutionProof}
\end{figure}

Let $P_{\mathrm{opt}}$ be a hamiltonian path which induces a
crossing-optimal HP-completion set for Case~1a. Recall that in
Case~1a the hamiltonian path is restricted to enter both nodes $t_i$
and $t_{i+1}$ from a node in the left side of $G$.
Figure~\ref{fig:dynProgSolutionProof} shows the structure of
$P_{\mathrm{opt}}$  for this case. Let's traverse $P_{\mathrm{opt}}$
backwards and justify the presence of the specific vertices on it.

The hamiltonian path $P_{\mathrm{opt}}$ of $G_{i+1}$ has to
terminate at $t_{i+1}$. Since it enters $t_{i+1}$ from a vertex on
the left side of $G$, it follows from
Lemma~\ref{lem:NoMixedSidesBook} that the vertices of the left side
of $o_{i+1}$, i.e., sequence $l(o_{i+1})$, precede $t_{i+1}$ on
$P_{\mathrm{opt}}$. Now observe that $t_i$ is the first vertex of
$l(o_{i+1})$. This is due to the fact that $G_i$ and $o_{i+1}$ share
an edge. Given that, in Case~1a, $P_{\mathrm{opt}}$ also enters
$t_i$ from a vertex $x$ on the left side of $G$, $t_i$ is preceded
on $P_{\mathrm{opt}}$ by a maximal path $P_1$ on $l(G_i)$ which ends
with vertex $x$. Let $u$ be the first vertex of $P_1$. Then, since
$P_1$ does not extend (backwards) before vertex $u$,
$P_{\mathrm{opt}}$ must continue backwards on a node on the right
side $r(G)$ of $G$. Then the vertex that precedes $u$ has to be
vertex $v$. To see this, observe that if any other vertex $y$  of
$r(G_{i+1})$ is placed after $v$ on $P_{\mathrm{opt}}$, then on
$P_{\mathrm{opt}}$ there is  a path from $v$ to $x$ which together
with the path from $x$ to $v$ on $r(G)$ form a cycle, a clear
contradiction since the HP-extended graph of $G$ must be acyclic.

By Lemma~\ref{lem:NoMixedSidesBook}, the vertices of $r(o_{i+1})$,
the last of which is $v$, have to precede vertex $u$ on
$P_{\mathrm{opt}}$. Let the first vertex of $r(o_{i+1})$ be vertex
$w$. Then, on $P_{\mathrm{opt}}$, $w$ must be preceded by $s_{i+1}$.
To see this, observe that if $w$ was preceded by  a vertex different
than $s_{i+1}$, then that vertex had to be located below edge
$(s_{i+1}, t_i)$. Then, edge $(s_{i+1}, t_i)$ (as well as the median
of $t_{i+1}$) would incur  at least two crossings, a clear
contradiction since  only one crossing per edge of $G$ is allowed.

Then, since $s_{i+1}$ is the last vertex of $r(o_i)$, by
Lemma~\ref{lem:NoMixedSidesBook} $r(o_i)$ has to precede $w$ on
$P_{\mathrm{opt}}$. Finally, the hamiltonian path $P_{\mathrm{opt}}$
has to be completed by a path $P$ from the source $s_1$ of $o_1$ to
the first vertex of $r(o_i)$.

Observe that the structure the crossing-optimal Hamiltonian path
$P_{\mathrm{opt}}$ for Case~1 is identical to the hamiltonian path
constructed in our proof. Also note that from $P_{\mathrm{opt}}$ we
can infer  hamiltonian paths for $o_{i+1}$ and $G_i$. These two
hamiltonian paths can be combined with our construction to yield a
hamiltonian path for $G_{i+1}$. Thus, if $P_{\mathrm{opt}}$ would
infer an HP-completion set with smaller number of crossings than our
construction, then it would also infer a hamiltonian path for $G_i$
with less than $c(G_i,L)$ crossings. This is impossible since, by
definition, $c(G_i,L)$ is the minimum number of crossings. So, the
constructed hamiltonian path infers an HP-completion set with a
minimum number of edge crossings.

Note that, there is not necessarily only one hamiltonian path that
yields an optimal solution. Since in our construction, we apply a
``minimum'' operator,  if both of the involved hamiltonian paths
yield the same number  of crossings, then we should have at least
two different but equivalent (wrt edge crossings) hamiltonian
paths.\qed
\end{proof}

Algorihtm~\ref{alg:AHPCCM} is a dynamic programming algorithm, based
on Lemmata~\ref{lem:optimalSubsolution}
and~\ref{lem:dynProgSolution}, which computes the minimum number of
edge crossings $c(G)$ resulting from the addition of a
crossing-optimal  HP-completion set  to an OT-$st$-digraph $G$. The
algorithm can be easily extended to also compute the corresponding
hamiltonian path $S(G)$.

\begin{algorithm2e}[tb]

\Input{An Outerplanar Triangulated $st$-digraph $G(V^l \cup V^r
\cup \{s,t\}, E)$.}

\Output{The minimum number of edge crossing $c(G)$ resulting from
the addition of a crossing-optimal  HP-completion set to a graph $G$
such that each edge of $G$ is crossed at most once.}

\caption{\textsc{Acyclic-HPC-CM($G$)}}

\label{alg:AHPCCM}

\begin{enumerate}
\item
Compute the $st$-polygon decomposition $\mathcal{D}(G)= \{ o_1,~
\ldots, o_\lambda \}$ of $G$;\;
\item For each element $o_i \in \mathcal{D}(G),~ 1 \leq i \leq
\lambda$, compute $c(o_i,L)$ and $c(o_i,R)$\\
\hspace*{.7cm}\textbf{if} $o_i$ is a free vertex, \textbf{then}
$c(o_i,L)=c(o_i,R)=0$. \\
\hspace*{.7cm}\textbf{if} $o_i$ is an $st$-polygon, \textbf{then}
$c(o_i,L)$ and $c(o_i,R)$ are computed based on\\
\hspace*{1.4cm}Lemma~\ref{lem:spineCrossingSTpolygon}.\\

\item \textbf{if} $o_1$ is a free vertex, \textbf{then} $c(G_1,L)=c(G_1,R)=0$;\\
\textbf{else}  $c(G_1,L)= c(o_1,L)$ and $ c(G_1,R)=c(o_1,R)$;
\item
 For $i=2\ldots \lambda,$ compute $c(G_i,L)$ and $c(G_i,R)$ as
 follows:\\
\hspace*{.7cm}  \textbf{if} $o_i$ is a free vertex, \textbf{then}\\
     \hspace*{1.4cm} $c(G_i,L)=c(G_i,R)=\min\{ c(G_{i-1},L), c(G_{i-1},R) \}$;

\hspace*{.7cm}  \textbf{else-if} $o_i$ is an $st$-polygon sharing \textbf{at most} one vertex with $G_{i-1}$, \textbf{then}\\
     \hspace*{1.4cm} $c(G_i,L)=\min\{ c(G_{i-1},L), c(G_{i-1},R)\} + c(o_1,L) $;
     \hspace*{1.4cm} $c(G_i,R)=\min\{ c(G_{i-1},L), c(G_{i-1},R)\}  + c(o_1,R)
     $;\\
\hspace*{.7cm}  \textbf{else} \{ $o_i$ is an $st$-polygon sharing \textbf{exactly} two vertices with $G_{i-1}$\}, \\

    \hspace*{1.4cm} \textbf{if} $t_{i-1} \in V^l$, \textbf{then} \\
        \hspace*{2.1cm}  $c(G_{i},L)=\min \{ c(G_{i-1},L)+c(o_{i},L)+1,~c(G_{i-1},R)+c(o_{i},L) \}
        $\\
        \hspace*{2.1cm}  $c(G_{i},R)=\min \{
        c(G_{i-1},L)+c(o_{i},R),~c(G_{i-1},R)+c(o_{i},R)\}$\\

    \hspace*{1.4cm} \textbf{else} \{ $t_{i-1} \in V^r$ \} \\
        \hspace*{2.1cm} $c(G_{i},L)=\min \{ c(G_{i-1},L)+c(o_{i},L),~c(G_{i-1},R)+c(o_{i},L)
\} $\\
        \hspace*{2.1cm}  $c(G_{i},R)=\min \{c(G_{i-1},L)+c(o_{i},R),~c(G_{i-1},R)+c(o_{i},R)+1
\} $\\
\item \textbf{return} $c(G)= \min\{ c(G_\lambda, L), c(G_\lambda,R) \}$
\end{enumerate}
\end{algorithm2e}

\begin{theorem}
\label{thm:optimalAcyclicHPCCM} Given an  $n$ node OT-$st$-digraph
$G$, a  crossing-optimal HP-completion set for $G$ with at most one
crossing per edge and the corresponding number of edge-crossings can
be computed in $O(n)$ time.
\end{theorem}

\begin{proof}
Algorithm~\ref{alg:AHPCCM} computes the number of crossings in an
acyclic HP-completion set. Note that it is easy to be extended so
that it computes the actual hamiltonian path (and, as a result, the
acyclic HP-completion set). To achieve this, we only need to store
in an auxiliary array the term that resulted to the minimum values
in Step~4 of the algorithm, together with the endpoints of the edge
that is added to the HP-completion set for each $st$-polygon in the
$st$-polygon decomposition $\mathcal{D}(G)= \{ o_1,~ \ldots,
o_\lambda \}$ of $G$. The correctness of the algorithm follows
immediately from Lemmata~\ref{lem:optimalSubsolution}
and~\ref{lem:dynProgSolution}.

From Lemma~\ref{lem:medianComputation} and
Theorem~\ref{thm:STpolygonDecomposition}, it follows that Step~1 of
the algorithm needs $O(n)$ time. The same hold for Step~2 (due to
Lemma~\ref{lem:spineCrossingSTpolygon}). Step~3 is an initialization
step that needs $O(1)$ time. Finally, Step~4 takes $O(\lambda)$
time. In total, the running time of  Algorithm~\ref{alg:AHPCCM} is
$O(n)$. Observe that $O(n)$ time is enough to also recover the
acyclic HP-completion set.\qed
\end{proof}

\section{Spine Crossing Minimization for Upward Topological 2-Page Book Embeddings of OT-$st$
Digraphs}

In this section,  we establish for the class of $st$-digrpahs an
equivalence (through a linear time transformation) between the
Acyclic-HPCCM problem and the problem of obtaining an upward
topological 2-page book embeddings with minimum number of spine
crossings and at most one spine crossing per edge. We exploit this
equivalence to develop an optimal (wrt spine crossings) book
embedding for  OT-$st$ digraphs.

\eat{
\begin{lemma} An optimal upward topological  2-page book embedding of a rhombus
has exactly one spine crossing.
\end{lemma}
\begin{proof}
Note that a rhombus does not have a hamiltonian path and that it has
an optimal HP-completion set consisting of a single edge. Then, the
lemma, follows as a consequence of
Theorem~\ref{thm:BEequivHPcompletionSet}. \qed
\end{proof}
}

\begin{theorem}
\label{thm:BEequivHPcompletionSet} Let $G=(V,E)$ be an $n$ node
$st$-digraph. $G$ has a crossing-optimal   HP-completion set $E_c$
with Hamiltonian path $P=(s=v_1, v_2, \ldots, v_n=t)$ such that the
corresponding optimal drawing $\Gamma(G^\prime)$ of $G^\prime=(V,E
\cup E_c)$  has $c$ crossings \textbf{if and only if} $~G$ has an
optimal (wrt the number of spine crossings) upward topological
2-page book embedding with $c$ spine crossings where the vertices
appear on the spine in the order $\Pi=(s=v_1, v_2, \ldots, v_n=t)$.
\end{theorem}

\begin{proof}

\begin{figure}[htb]
    \begin{minipage}{\textwidth}
    \centering
    \includegraphics[width=.75\textwidth]{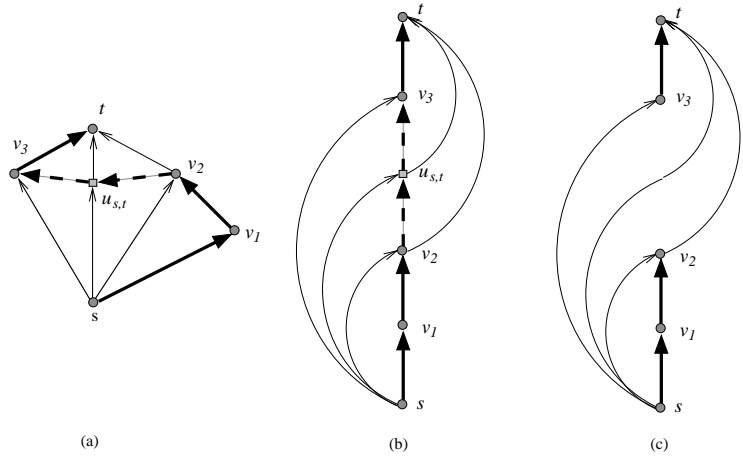}
    \caption{(a) A drawing of an HP-extended digraph for an  $st$-digraph $G$.
     The dotted segments correspond to the single edge $(v_2, v_3)$ of the HP-completion set for $G$.
     (b)An upward topological 2-page book embedding of $G_c$ with
     its
     vertices   placed on the spine in the order they appear
     on a hamiltonian path of $G_c$.
     (c)An upward topological 2-page book embedding of $G$. }
    \label{fig:HPcompletionExample}
  \end{minipage}
\end{figure}

We  show how to obtain from an HP-completion set with $c$ edge
crossings  an upward topological 2-page book embedding with $c$
spine crossings and vice versa. From this is follows that   a
crossing-optimal  HP-completion set for $G$ with $c$ edge crossing
corresponds to an optimal upward
topological 2-page book embedding  with the same number of spine crossings.\\
``$\Rightarrow$''$~~~$ We assume that we have an HP-completion set
$E_c$  that satisfies the conditions stated in the theorem. Let
$\Gamma(G^\prime)$ of $G^\prime=(V,E \cup E_c)$ be the corresponding
 drawing that has $c$ crossings and let $G_c=(V\cup V_c,
E^\prime \cup E_c^\prime)$ be the acyclic HP-extended digraph of $G$
wrt $\Gamma(G^\prime)$.   $V_c$ is the set of new vertices placed at
each edge crossing. $E^\prime$  and $E_c^\prime$ are the edge sets
resulting from $E$ and $E_c$,
 respectively, after splitting their edges
 involved in crossings and maintaining their orientation
 (see Figure~\ref{fig:HPcompletionExample}(a)). Note that $G_c$ is also an $st$-planar digraph.

 Observe that in $\Gamma(G^\prime)$ we have no crossing involving two edges of $G$.
 If this was the
 case, then $\Gamma(G^\prime)$ would not preserve $G$.
 Similarly, in $\Gamma(G^\prime)$ we have no crossing involving two edges of the HP-completion set $E_c$.
 If this was the
 case, then $G_c$ would contain a cycle.

 The hamiltonian path $P$ on $G^\prime$ induces a hamiltonian path
 $P_c$ on the HP-extended digraph $G_c$. This is due to the facts that all
 edges of $E_c$ are used in hamiltonian path $P$ and all vertices of
 $V_c$ correspond to crossings involving edges of $E_c$. We  use
 the hamiltonian path $P_c$ to construct an upward topological
 2-page book embedding for graph $G$ with exactly $c$ spine
 crossings. We place the vertices of $G_c$ on the spine in the order
 of hamiltonian path $P_c$, with vertex $s=v_1$ being the lowest.
 Since the HP-extended digraph $G_c$ is a planar $st$-digraph with vertices $s$ and $t$ on the external face,
 each edge of $G_c$
 appears either to the left or to the right of the hamiltonian path
 $P_c$.
 We place the  edges of $G_c$  on the left (resp. right) page of the book embedding if
 they appear to the left (resp. right) of path $P_c$. The edges of
 $P_c$ are drawn on the spine (see Figure~\ref{fig:HPcompletionExample}(b)).
 Later on they can be moved to any of
 the two book pages.

 Note that all edges of $E_c$ appear on the spine. Consider any vertex  $v_c \in V_c$.
 Since $v_c$ corresponds to a crossing between an edge of $E$ and an edge of
 $E_c$, and the edges of $E_c^\prime$ incident to it have been drawn
 on the spine, the two remaining edges of $E^\prime$ correspond to (better, they are parts of) an edge $e \in E$
 and  drawn on
 different pages of the book.   By removing vertex $v_c$ and merging
 its two incident edges of $E^\prime$ we create a crossing of edge
 $e$ with the spine. Thus, the constructed book embedding has as
 many spine crossings as the number of edge crossings of
 HP-completed graph $G^\prime$ (see Figure~\ref{fig:HPcompletionExample}(c)).

 It remains to show that the
 constructed book embedding is upward. It is sufficient to show that the constructed book
 embedding of $G_c$ is upward. For the sake of contradiction, assume
 that there exists a downward edge $(u,w) \in E_c^\prime$. By the construction, the fact
 that $w$ is drawn below $u$ on the spine implies that there is a
 path in $G_c$ from $w$ to $u$. This path, together with edge
 $(u,w)$ forms a cycle in $G_c$, a clear contradiction since $G_c$
 is acyclic.

``$\Leftarrow$''$~~~$ Assume that we have an upward 2-page
topological book embedding of $st$-digraph $G$ with $c$ spine
crossings where the vertices appear on the spine in the order
$\Pi=(s=v_1, v_2, \ldots, v_n=t)$. Then, we construct an
HP-completion set $E_c$ for $G$ that satisfies the condition of the
theorem as follows: $E_c = \{ (v_i, v_{i+1}) ~|~ 1 \leq i <n
\mbox{~and~} (v_i, v_{i+1}) \not\in E \}$, that is, $E_c$ contains
an edge for each consecutive pair of vertices of the spine that (the
edge) was not present in $G$. By adding/drawing these edges on the
spine of the book embedding we get a drawing  $\Gamma(G^\prime)$ of
$G^\prime=(V,E \cup E_c)$ that has $c$ edge crossings. This is due
to the fact that all spine crossing of the book embedding are
located, (i) at points of the spine above vertex $s$ and below
vertex $t$, and (ii) at points of the spine between consecutive
vertices that are not connected by an edge. By inserting at  each
crossing of $\Gamma(G^\prime)$  a new vertex and by splitting the
edges involved in the crossing while maintaining their orientation,
we get an HP-extended digraph $G_c$. It remains to show that $G_c$
is acyclic. For the sake of contradiction, assume that $G_c$
contains a cycle. Then, since graph $G$ is acyclic, each cycle of
$G_c$ must contain a segment resulting from the splitting of an edge
in $E_c$. Given that in $\Gamma(G^\prime)$ all vertices appear on
the spine and all edges of $E_c$ are drawn upward, there must be a
segment of an edge of $G$ that is downward in order to close the
cycle. Since, by construction, the book embedding of $G$ is a
sub-drawing of $\Gamma(G^\prime)$,  one of its edges (or just a
segment of it) is downward. This is a clear contradiction since we
assume that the topological 2-page book embedding of $G$ is upward.
\qed
\end{proof}

\begin{theorem}
\label{thm:optimalAcyclicHPCCM} Given an  $n$ node OT-$st$-digraph
$G$, an upward 2-page topological book embedding for $G$ with
minimum number of spine crossings and at most one spine crossing per
edge and the corresponding number of edge-crossings can be computed
in $O(n)$ time.
\end{theorem}

\begin{proof}
By Theorem~\ref{thm:BEequivHPcompletionSet} we know that by solving
the Acyclic-HPCCM problem on $G$, we can deduce the wanted upward
book embedding. By Theorem~\ref{thm:optimalAcyclicHPCCM}, the
Acyclic-HPCCM problem can be solved in $O(n)$ time.\qed
\end{proof}


\newpage
\section{Conclusions - Open Problems}
\begin{figure}[htb]
    \begin{minipage}{\textwidth}
    \centering
    \includegraphics[width=0.3\textwidth]{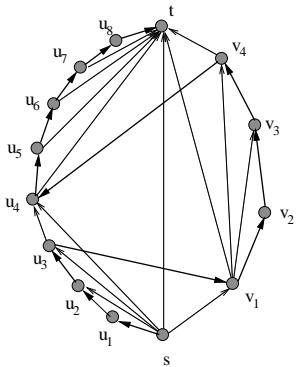}
    \caption{The paths $\langle s,u_1,\dots,u_8,v_1,\dots,v_4,t\rangle$ and
    $\langle s,v_1,\dots,v_4,u_1,\dots,u_8,t\rangle$ have 5 crossings each, with one crossing per edge.
    Path
    $\langle s,u_1,u_2,u_3,v_1,\dots,v_4,u_4,\dots,u_8,t\rangle$ has 4 crossings,
    two of which involve edge $(s,t)$.}
    \label{fig:cntrexmpl}
  \end{minipage}
\end{figure}

We have studied the problem of Acyclic-HPCCM and we have presented a
linear time algorithm that computes a crossing-optimal Acyclic
HP-completion set for an OT $st$-digraphs $G$ with at most one
crossing per edge of $G$. Future research include the study of the
Acyclic-HPCCM on the larger class of $st$-digraphs, as well as
relaxing the requirement for $G$ to be triangulated. Another
interesting research direction is to derive HP-completion sets for
the (Acyclic-) HPCCM problem that can have any number of crossing
with the edges of graph $G$. Figure~{\ref{fig:cntrexmpl}} shows an
$st$-polygon which becomes hamiltonian by adding one of the
following completion sets: $A=\{(u_8, v_1)\}$, $B=\{(v_4, u_1)\}$ or
$C=\{(u_3,v_1), (v_4,u_4)\}$. Sets $A$ and $B$ creates 5 crossings
with one crossing per edge of $G$ while, set $C$ creates 4 crossings
with at most 2 crossings per edge of $G$.

\bibliographystyle{abbrv}

\end{document}